\documentclass[12pt]{iopart}
\usepackage{graphics}

\newcommand{\lnc}{LaNiC$_2$}
\newcommand{\lpb}{Li$_2$Pt$_3$B}
\newcommand{\cps}{CePt$_3$Si}
\newcommand{\cis}{CeIrSi$_3$}
\newcommand{\crs}{CeRhSi$_3$}
\newcommand{\ccg}{CeCoGe$_3$}

\begin{document}

\title[]{Nodal gap structure in the noncentrosymmetric
superconductor LaNiC$_2$ from magnetic-penetration-depth
measurements}

\author{I Bonalde$^1$, R L Ribeiro$^1$\footnote{Present address: Laboratoire National des Champs Magn\'{e}tiques Intenses,
INSA UPS UJF CNRS, UPR 3228, Universit\'{e} de Toulouse, 143
av. de Rangueil, 31400 Toulouse, France.}, K J Syu$^2$, H H
Sung$^2$ and W H Lee$^2$}

\address{$^1$ Centro de F\'{\i}sica, Instituto Venezolano de
Investigaciones Cient\'{\i}ficas, Apartado 20632, Caracas
1020-A, Venezuela}

\address{$^2$ Department of Physics,
National Chung Cheng University, Ming-Hsiung, Chia-Yi 62199,
Taiwan, ROC}

\begin{abstract}
We report measurements of the temperature dependence of the
magnetic penetration depth in different quality polycrystalline
samples of noncentrosymmetric LaNiC$_2$ down to 0.05 K. This
compound has no magnetic phases and breaks time-reversal
symmetry. In our highest quality sample we observe a $T^2$
dependence below $0.4T_c$ indicative of nodes in the energy
gap. We argue that previous results suggesting conventional
$s$-wave behavior may have been affected by magnetic
impurities.
\end{abstract}



\ead{ijbonalde@gmail.com}

\maketitle

\section{Introduction}

Noncentrosymmetric superconductors have gained interest since
the discovery of superconductivity in \cps\ \cite{bauer}. The
absence of inversion symmetry leads to the indistinguishability
of spin-singlet and spin-triplet states and to the appearance
of an antisymmetric spin-orbit coupling (ASOC) that splits the
electron bands by lifting the spin degeneracy. Thus unusual
superconducting properties have been expected in all these
materials, though to date such properties have been observed
only in a small group: \cps\ \cite{mine11,mine13},
 \lpb\ \cite{yuan2,nishiyama}, \cis,\ \crs,\ \ccg,\ and
 CeIrGe$_3$ \cite{kimura07,measson09,mine17}. With the notable
 exception of \lpb,\ in this group superconductivity appears
 inside an antiferromagnetic order and mostly around a quantum critical
 point. This suggests that the unusual behaviors may originate from
the interplay of antiferromagnetic interaction and ASOC
\cite{fujimoto2,yanase1}.

Nonmagnetic \lnc\ is an intriguing noncentrosymmetric
superconductor~\cite{whlee1} in which both unconventional and
conventional behaviors have been uncovered. Unconventional
characteristics were found in early specific-heat measurements
\cite{whlee1} that showed a low-temperature $T^3$ dependence
expected when the energy gap has nodes and in recent
muon-spin-relaxation results \cite{hillier} that suggested the
lack of time-reversal symmetry (TRS). NQR-$1/T_1$
\cite{iwamoto} and recent specific-heat \cite{pecharsky} data
were interpreted in terms of conventional BCS
superconductivity. On the other hand, theoretical approaches
\cite{fujimoto2,yanase1,subedi} imply that \lnc\ should behave
like a conventional superconductor. The fact that \lnc\ is
nonmagnetic and has a relatively strong ASOC~\cite{hase} makes
the determination of its pairing symmetry important for the
understanding of the physics of noncentrosymmetric
superconductors. Moreover, the symmetry of the order parameter
of \lnc\ may be relevant for superconductivity in general,
because a) in the absence of inversion symmetry only some
spin-triplet states are allowed \cite{frigeri}, b) in the
absence of TRS spin-singlet states are forbidden, and c) TRS is
required by the presence of ASOC \cite{frigeri}.

The conflicting experimental results in \lnc\ may have been
caused by the poor quality of the samples and the not-so-low
temperatures of the experiments. It is known that in order to
determine energy gap structures from thermodynamic and
transport properties temperatures below $0.3T_c$ are needed.
Here, we present measurements of the magnetic penetration depth
$\lambda(T)$ in different quality samples of \lnc\ down to 50
mK ($\sim 0.017T_c$). Penetration depth is highly recognized as
a unique probe of the structure and symmetry of the
superconducting order parameter. In our highest quality sample
we found that $\lambda(T)\propto T^2$ as $T\longrightarrow 0$,
which suggests the existence of nodes in the energy gap.

\section{Experimental methods}
%
\begin{figure}
\begin{center}
\scalebox{0.43}{\includegraphics{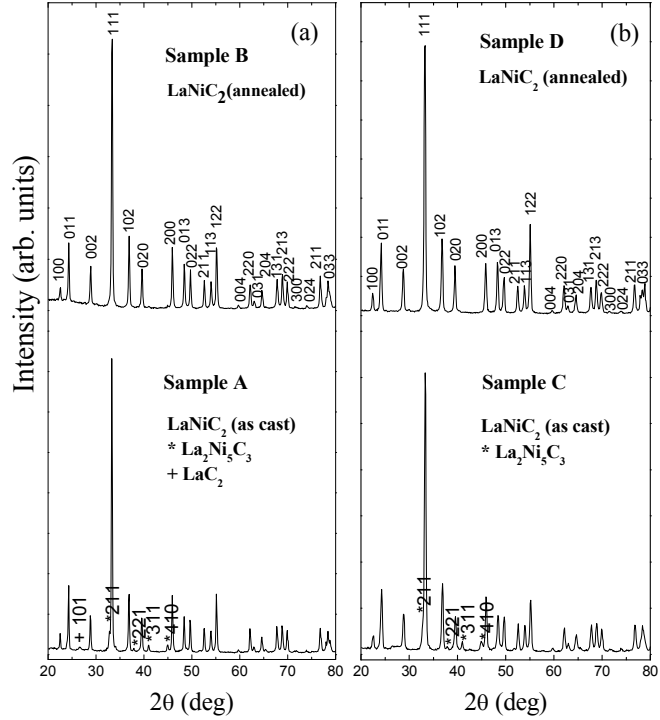}}
\caption{Room-temperature powder x-ray diffraction patterns for
our LaNiC$_2$ samples prepared with (a) 99.9\% Ni and (b)
99.995\% Ni.} \label{xray}
\end{center}
\end{figure}
%
%

\lnc\ crystallizes in the orthorhombic CeNiC$_2$-type structure
with space group $Amm2$ \cite{whlee1}. We studied four
different polycrystalline samples (labeled A, B, C, and D)
prepared by arc melting suited amounts of La (99.9\%, Ames
Lab), C (99.9995\%, Alpha Aesar) and Ni in two degrees of
purity. Samples A and B were prepared with 99.9\% Ni (40 ppm of
Fe, Alpha Aesar) and samples C and D with 99.995\% Ni (13 ppm
of Fe, Alpha Aesar). Samples B and D were then sealed under
argon in quartz tubes and annealed at 1273 $^\circ$C for 10
days, and finally water quenched to room temperature. To check
for impurity phases in the samples, we performed powder x-ray
diffraction measurements using a microcomputer controlled MXP3
diffractometer with graphite monochromated Cu $K\alpha$
radiation. Annealed samples B and D displayed sharp diffraction
lines and no additional reflections (see upper spectra in
Fig.~\ref{xray}), which is taken as evidence of single phases.
In contrast, as-cast samples A and C showed additional
reflection lines (see lower spectra in Fig.~\ref{xray})
corresponding to the secondary phases LaC$_2$ and
La$_2$Ni$_5$C$_3$. In samples A, B and C the transition
temperature $T_c \approx 3.6$ K, whereas in sample D $T_c
\approx 3$ K.

Penetration-depth measurements were carried out utilizing a 14
MHz tunnel diode oscillator. The deviation of the penetration
depth $\lambda(T)$ from its value at the lowest measured
temperature, $\Delta\lambda(T)=\lambda(T)-\lambda$(0.05 K), was
obtained up to $T \sim 0.99T_c$ from the corresponding change
in the measured resonance frequency $\Delta f(T)$: $\Delta f(T)
= G \Delta \lambda(T)$. Here $G$ is a constant factor that
depends on the sample and coil geometries and that includes the
demagnetizing factor of the sample. We estimated $G$ by
measuring a sample of known behavior and of the same dimensions
as the test sample~\cite{mine10}. To within this calibration
factor, $\Delta \lambda(T)$ is raw data.

We note here that the low-temperature dependence of
$\lambda(T)$ does not seem to be affected by the sample type
(single crystal, polycrystal, etc.) \cite{mine13,mine7}. In any
case, intergrain or proximity effects are not expected to be
relevant in the present results, because the measuring magnetic
field is very small (about 5 mOe) \cite{goldfarb}. For
comparison, we measured a pure (99.999\%) polycrystalline
sample of the $s$-wave superconductor In ($T_c=3.4$ K).
Intergranular susceptibility normally saturates at low
temperatures.

\section{Results and Discussion}

%
\begin{figure}
\begin{center}
\scalebox{0.45}{\includegraphics{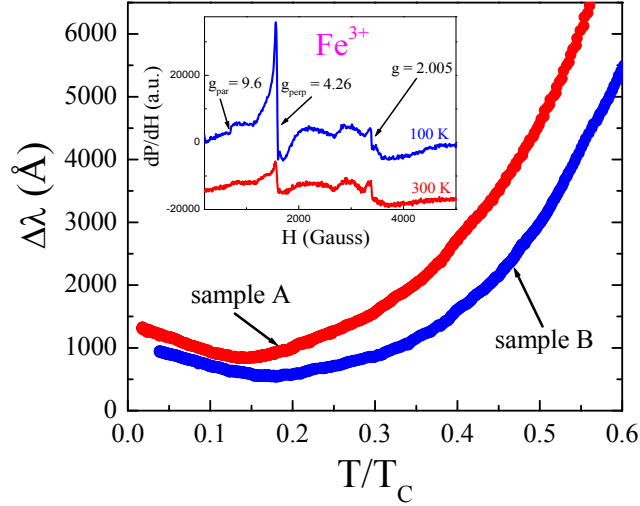}}
\caption{Low-temperature $\Delta\lambda(T)$ of samples A and B.
The upturns below $0.2T_c$ are due to a contribution coming
from Fe impurities in the samples. Inset: EPR spectra at
different temperatures for sample B that clearly indicate the
presence of interstitial Fe$^{3+}$. \textrm{g}$_{par}$ and
\textrm{g}$_{perp}$ are the parallel and perpendicular
 components of the \textrm{g}-factor, respectively.}
\label{lambdaAB}
\end{center}
\end{figure}
%
%

Figure~\ref{lambdaAB} displays the low-temperature region of
$\Delta\lambda(T)$ of samples A and B. Both curves have an
upturn below $0.2T_c$ that in polycrystalline samples is
usually caused by the competition between the superconducting
screening effect and the magnetic permeability (whose thermal
response is due to magnetic-impurity effects)~\cite{cooper}. To
check for the presence of magnetic impurities in samples A and
B, we performed EPR spectroscopy at 100 and 300 K. In both
samples the spectra indicate the presence of interstitial
Fe$^{3+}$ in the orthorhombic structure of \lnc\ (see the
spectra of sample B in the inset to Fig.~\ref{lambdaAB}). Thus,
the upturns in the penetration-depth data of these samples
could indeed be produced by magnetic impurities. Previous
measurements in \lnc\ samples prepared with 99.9\% Ni, i.e.
similar to samples A and B, were carried out with probes
sensitive to magnetic impurities and were thus most likely
affected by the Fe$^{3+}$ impurities. Notably, the results of
all these measurements -$1/T_1$-NQR \cite{iwamoto},
magnetization and heat capacity \cite{pecharsky}- point to
conventional BCS behavior.

%
%
\begin{figure}
\begin{center}
\scalebox{0.5}{\includegraphics{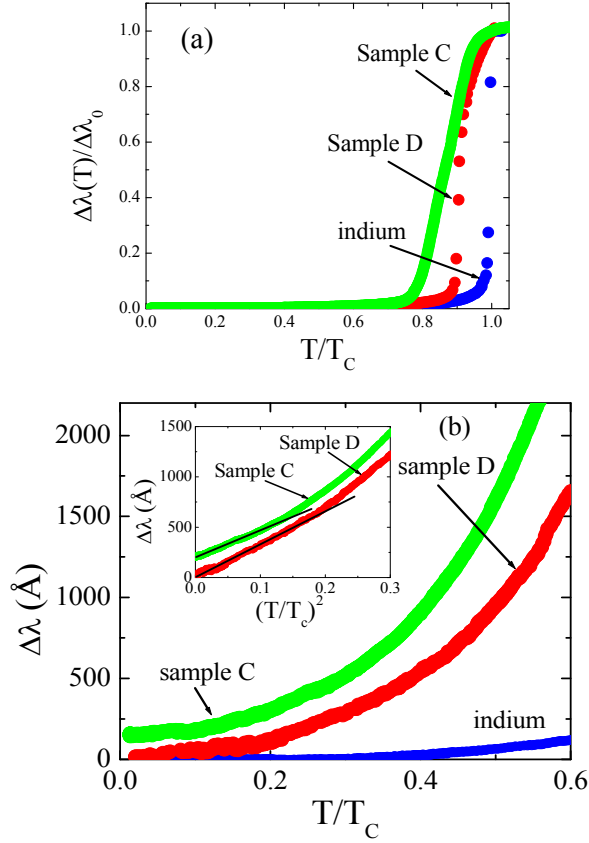}} \caption{(a)
Normalized $\Delta \lambda(T)$ against $T/T_c$ of samples C and
D and of conventional indium. (b) Closeup of the
low-temperature region showing that \lnc\ data follow a power
law instead of an exponential behavior (as the indium data do).
Inset: Low-temperature data plotted as a function of
$(T/T_c)^2$, where $\Delta\lambda(T)\propto T^2$ is observed
below $0.4T_c$ in sample D.}\label{lambdalowT}
\end{center}
\end{figure}
%
%
Figure~\ref{lambdalowT}a shows
$\Delta\lambda(T)/\Delta\lambda_0$ of sample C, sample D, and
indium. The data were scaled by the corresponding total
penetration depth shift $\Delta\lambda_0$ so as to compare
them. The superconducting transitions in these \lnc\ samples
are quite broad compared with the one in indium. Similar wide
transitions are seen in samples A and B. The broadness is
different from sample to sample which suggests that the wide
transitions in \lnc\ are due to defects or high inhomogeneity.

Figure~\ref{lambdalowT}b depicts the low-temperature region of
$\Delta\lambda(T)$ of samples C and D along with that of
indium. No upturn is observed in these samples of \lnc,\
prepared with the purest Ni (99.995\%), which could imply that
the level of Fe impurities is so small as to affect
significantly the superconducting properties. In samples C and
D the low-temperature magnetic penetration depth goes as $T^n$,
with $n\sim 2-2.4$, as opposed to the distinctly different
exponential behavior observed in the sample of the fully-gapped
superconductor indium (see figure). Inset to
Fig.~\ref{lambdalowT}b displays $\Delta\lambda(T)$ versus
$(T/T_c)^2$ for samples C and D in the temperature region
$T\leq0.55T_c$. Clearly, $\Delta\lambda(T)\propto T^2$ up to
about $0.4T_c$ in sample D. Since the annealed sample D has the
sharpest transition and the smallest exponent $n$ in a larger
temperature region, we consider the result in this sample to be
the closest to the true superconducting behavior in \lnc.\

A $T^2$ dependence of $\lambda(T)$ at the \textit{lowest}
temperatures would only result if the superconducting energy
gap has nodes or conventional gapless superconductivity occurs
(due to strong scattering effects) \cite{gross}. We note that
$\Delta\lambda(T)\propto T^2$ is the limiting behavior of
gapless superconductivity under impurity scattering; thus the
fact that we see a higher exponent $n$ in somewhat lower
quality samples would rule out gapless superconductivity.
Moreover, gapless superconductivity leads to linear temperature
behaviors in the electronic specific heat $C_e$ and the NMR
relaxation time $1/T_1$ \cite{sigrist2}. In previous works it
was found that $C_e \propto T^3$ in a sample of the same
quality as D \cite{whlee1} and that $C_e$ and $1/T_1$ go
exponentially with temperature in lower quality samples
\cite{iwamoto,pecharsky}. Thus, we argue here that the $T^2$
dependence of $\lambda(T)$ suggests the existence of nodes in
the gap function and, therefore, that superconductivity is
unconventional in \lnc.\

We now discuss the node origin of $\Delta\lambda(T)\propto
T^2$. Since we used polycrystalline samples both the in-plane
and the out-of-plane component of $\lambda(T)$ contributed to
our measured signals. Unfortunately, the available information
for \lnc\ is not sufficient to estimate its anisotropy that
would allow us to weight the contribution from each
penetration-depth component. A $T^2$ response can appear in the
out-of-plane component $\Delta\lambda_c(T)$ if point nodes are
present. This would be in agreement with the result $C_e
\propto T^3$ found earlier \cite{whlee1} that implies the
existence of point nodes in the gap. We notice that the
TRS-breaking states proposed under symmetry arguments
\cite{hillier,quintanilla} as the candidates for the order
parameter of \lnc\ do not posses point nodes.

A $T^2$ dependence could also be displayed in the in-plane
component $\Delta\lambda_{ab}(T)$ because of symmetry-imposed
line nodes in the presence of unitary scattering. In this case,
the quadratic response would be the result of a crossover from
the linear behavior of $\Delta\lambda_{ab}(T)$ expected for a
clean and pure material \cite{HiGo}. In \lnc\
impurities/defects may act as strong scatterers, since they
change the low-$T$ behavior of $\lambda(T)$. The crossover
temperature $T^*\sim\Delta(0)\sqrt{(T_{c0}-T_c)/T_{c0}}$ (with
$T_{c0}$ the impurity-free $T_c$) \cite{HiGo}, may apply up to
a correction factor of order one in noncentrosymmetric
superconductors \cite{mine13,frigeri2,mineev2}. In \lnc\
$T_{c0}$ is unknown and a fit to the crossover expression
$\Delta\lambda(T)= aT^2/(T^*+T)$ yielded $T^*>T_c$, consistent
with the fact that $\Delta\lambda(T) \propto T^2$ up to high
temperatures. This result indicates that the impurity-free
critical temperature should be appreciably higher than 3-3.6 K,
and there is not evidence of it. Thus, it is unclear whether
line nodes affected by impurities/defects may cause the $T^2$
dependence. It is worth mentioning that in the
noncentrosymmetric superconductor \cps\ the linear behavior of
$\lambda(T)$ at low temperatures was found to be robust against
impurities/defects \cite{mine13}.

Even though the present results indicate that the energy gap of
\lnc\ has nodes, we believe that the type of node would be only
elucidated when pure single crystals become available.

%
%
\begin{figure}
\begin{center}
\scalebox{0.45}{\includegraphics{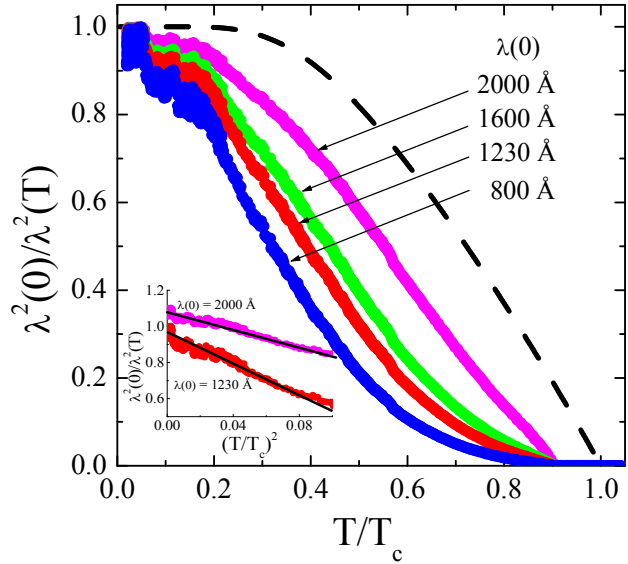}} \caption{Superfluid
density $\rho(T)\propto \lambda^2(0)/\lambda^2(T)$ of sample D
and the numerical data of a conventional $s$-wave local model
(dashed line). Due to the uncertainty in $\lambda(0)$,
$\rho(T)$ was plotted for several values of $\lambda(0)$.
Inset: Low-temperature $\rho(T)$ as a function of $(T/T_c)^2$
for $\lambda(0)= 1230$ $\AA$ and $\lambda(0)=2000$ $\AA$. In
both cases the data follow a $T^2$ law. The solid lines are
guides to the eye. }\label{supdennum}
\end{center}
\end{figure}
%

In Fig.~\ref{supdennum} we compare the superfluid density
$\rho(T)\propto \lambda^2(0)/\lambda^2(T)$ of sample D with the
numerical data of a conventional $s$-wave local model. We
estimated $\lambda(0)\sim 1230$ $\AA$ from $\gamma_n\approx 7$
mJ/mol$\cdot$K$^2$ and $H_{c2}(0)\sim 1250$ Oe
\cite{whlee1,iwamoto}. Since $\lambda(0)$ has not been
experimentally obtained, in Fig.~\ref{supdennum} we plotted
$\rho(T)$ for several values of $\lambda(0)$. In all cases the
disagreement between theory and experiment is evident, which
supports the argument that \lnc\ does not behave as a
conventional $s$-wave superconductor. The inset to this figure
shows that at low temperatures the superfluid density goes as
$T^2$ independently of the value of $\lambda(0)$, in accordance
with the penetration-depth data.

The strong suppression of the superfluid density at high
temperatures is similar to that found in other
noncentrosymmetric superconductors with and without nodes
\cite{mine9,mine14}. Moreover, in the well-established $d$-wave
superconductors $\kappa$-(ET)$_2$Cu[N(CN)$_2$]Br and
$\kappa$-(ET)$_2$Cu(NCS)$_2$ the superfluid density shows
upward curvature~\cite{mine1}. Recently, however, it has been
widely considered that an upward curvature in the superfluid
density is a signature of two-gap superconductivity. We argue
that, regardless of the curvature of the superfluid density, in
the \emph{true} low-temperature limit two-isotropic-gap
superconductors display a temperature-independent behavior,
whereas superconductors with nodes in the energy gap show a
power-law response. That is, as $T\longrightarrow0$ a
temperature-independent behavior of $\lambda(T)$ or $\rho(T)$
implies a nodeless energy gap, while a power-law response is a
very strong evidence of nodes.

Our results indicate that (1) previous measurements in
low-quality samples that suggest conventional $s$-wave
behaviors may have been affected by magnetic impurities and (2)
in higher quality samples (like our sample D) the
superconducting properties of \lnc\ are characterized by an
energy gap with nodes. Thus, \lnc\ is only the second
noncentrosymmetric superconductor without magnetic phases or
strong electron correlations found to have nodes in the energy
gap. Until now \lpb,\ which also has a relatively strong ASOC,
was the only example of such a superconductor. Based on the
results in \cps,\ \cis,\ and \crs,\ most recent models for
noncentrosymmetric superconductors point out that the existence
of (accidental) nodes requires the presence of both an
antiferromagnetic interaction and a sizable ASOC
\cite{fujimoto2,yanase1}. Our results in \lnc\ suggest that an
antiferromagnetic coupling may not be required for the
existence of nodes in superconductors without inversion
symmetry.

\section{Conclusions}

In summary, we reported measurements of the magnetic
penetration depth in polycrystalline samples of
noncentrosymmetric \lnc.\ The very-low-temperature penetration
depth of our highest quality sample followed a $T^2$ law that
strongly suggest nodes in the energy gap. We believe that
previous results indicating conventional $s$-wave behaviors
were most probably affected by magnetic impurities.

\ack We would like to thank Dr Marisel D\'{i}az at Instituto
Venezolano de Investigaciones Cient\'{i}ficas (IVIC) for the
EPR spectra. This work was supported by IVIC Project No. 441.

\section*{References}

\end{document}